\documentclass[letter]{aa} 

\usepackage{graphicx}
\usepackage{txfonts}
\usepackage{color}
\definecolor{gray}{rgb}{0.5,0.6,0.7}
\definecolor{raspberry}{rgb}{0.7,0.2,0.4}
\definecolor{emerald}{rgb}{0,0.61,0.47}

\begin{document}

   \title{Tidal origin of dark matter-free dwarf galaxies in the NGC 1052 group}

   \titlerunning{Dark matter-free dwarfs around NGC 1052}

   \author{F. Hammer\inst{1} and Y. B. Yang\inst{1} 
          }
\authorrunning{F. Hammer and Y. B. Yang}
   \institute{LIRA, Observatoire de Paris, PSL University, CNRS, 
              \email{francois.hammer@obspm.fr}             }
   \date{Received 30 March, 2026; Accepted 06 August, 2026}

 
  \abstract
   {The discovery of dark matter-free (DM-free) dwarf galaxies in the NGC 1052 neighborhood has had a considerable impact on modern cosmology. The galaxies have been explained through a dwarf--dwarf head-on collision, a rare event. We find that they could alternatively be associated with a head-on 1:1 merger after it was tuned to generate the E4 morphology of NGC 1052. Our simulations show that such mergers produce long-lived tidal features, are associated with the remnant galaxy, and are in the form of large tidal tails, including tidal dwarf galaxies (TDGs). We emphasise that such tidal features are predicted by the hierarchical scenario in which massive galaxies are formed by galaxy mergers. The latter can reproduce both the tidal features in the NGC1052 outskirts and the observed dwarf galaxies. The simulated TDGs have sizes similar to those observed, while they are ten times smaller in the bullet dwarf scenario. However, we cannot reproduce the luminous globular cluster systems due to resolution limitations. Resolving the radial distance between the DM-free dwarfs is necessary to identify the scenario of their formation. We suggest that there should be many other examples of DM-free dwarf galaxies in the neighbourhood of local massive galaxies and galaxy groups.}

   \keywords{Elliptical galaxies --
                dark matter -- merger --
                dynamics
              }

   \maketitle
%
\section{Introduction}
The discovery of dark matter (DM) deficient dwarf galaxies in the NGC 1052 group has attracted considerable attention and generated many debates over the past 8 years \citep{vanDokkum2018,vanDokkum2019}. For example, \citet{Monelli2019} argued that the mass based on dynamics of surrounding globular clusters (GCs) of NGC1052-DF2 could have been underestimated due to a distance overestimate. However, a more accurate distance  based on the tip of the red giant branch (TRGB) stars has been obtained \citep{Shen2021}, confirming the distance of two DM-deficient dwarfs, DF2 and DF4, at $\sim$ 20 Mpc (see also \citealt{Tang2026}). 

Galaxies are known to grow within a dominant DM halo in the current cosmological frame. To reconcile the latter with the absence of DM in some dwarf galaxies, \citet{Silk2019} proposed the bullet dwarf collision scenario in which two dwarfs experience a collision sufficiently energetic to generate DM-free galaxies. The scenario has been explored by several numerical simulations \citep[and references therein]{Lee2024}.  \citet{Lee2024} showed from comparison with cosmological simulations that such an event could occur despite being a rare event in the field. Their model approximately reproduces the remarkable alignment of at least five additional dwarfs with the line drawn from NGC1054-DF2 to DF4 , both on sky \citep{vanDokkum2022}  and in radial velocity \citep{Keim2025} projections. 

The alignment may also suggest a TDG origin since many tidal tails formed during galaxy merger episodes show an extended linear appearance. The NGC 1052 group is made up of a core including four massive galaxies within a projected radius of 150 kpc, which favours a rich merger history. NGC 1052 is a massive elliptical galaxy that shows signs of tidal perturbations, such as a narrow and linear stellar stream that points near NGC1052-DF2 \citep{Muller2019}. 
Using numerical simulations, \citet{Bournaud2006} showed that tidal satellites should be found more frequently around early-type galaxies.  
The formation of an elliptical galaxy has been modelled by, for example, \cite{Wang2020} to reproduce the properties of Cen A. Both Cen A and NGC 1052 have Seyfert 2 type active nuclei and massive black holes \citep{Wang2020,Koss2022}, whose formation and growth are likely related to mergers \citep{Hopkins2006}.

In this paper we investigate whether the chain of DM-deficient dwarfs could be associated with a tidal tail, which would have been formed during a merger similar to that described by  \cite{Wang2020}. The goal is not to fully reproduce the observations of the NGC 1052 field but to verify whether this hypothesis may be compelling in comparison with the bullet dwarf model described by \citet{Lee2024} and by \citet{Keim2025}. In Sect.~\ref{sec:merger} we describe the mechanism and associated parameters necessary to form an elliptical galaxy such as Cen A and NGC 1052. Section~\ref{sec:reproduction} shows how such a model may generate tidal tails and associated tidal dwarfs, and we also compare how realistically it may reproduce the observations. In Sect.~\ref{sec:discussion} we compare and discuss the strengths and weaknesses of the two competitive models, the result of which is summarised in Section~\ref{sec:conclusion}.
   
   \begin{figure}
   \centering
\includegraphics[width=9cm]{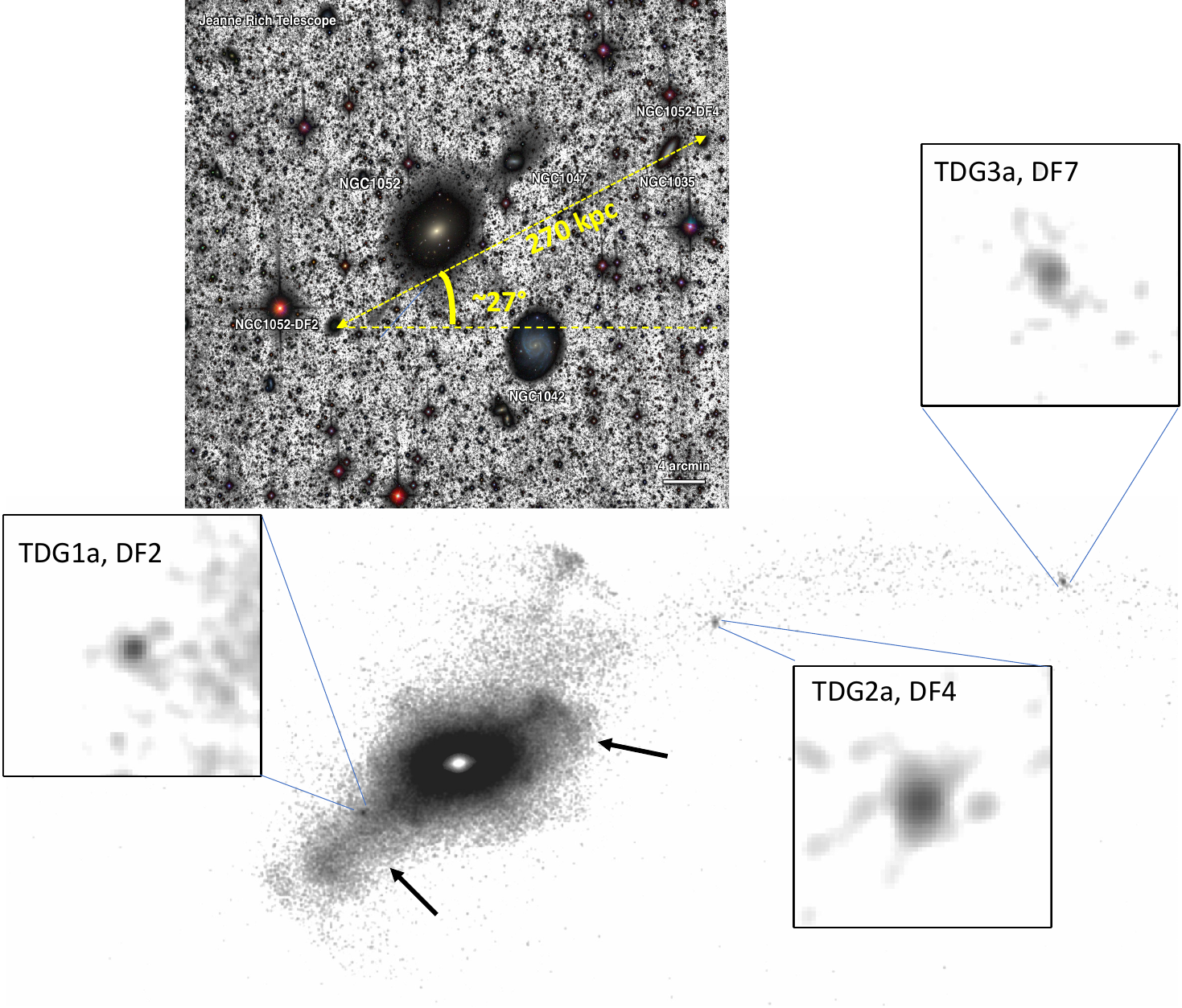}
      \caption{Comparison between the deep observations by \citet[see top-left image]{Muller2019} and the \citet{Wang2020} adapted simulations for stellar particles (bottom panel). In the insert panels, three dwarfs are zoomed in on that may resemble DF2, DF4, and DF7 (out of the observed field of view); their properties are given in Table~\ref{tab:TDGs}. The dwarf spatial alignment shows many similarities with observations. NGC 1052 is reproduced as an E4 galaxy (see the image in the centre of the halo) with tidal features (see arrows)  observed by \citet[see top-left image]{Muller2019}, although the halo orientation is offset by 45 degrees. The projected distance (270 kpc) of DF2--DF4 is indicated for both observations. Scaled simulations (in yellow) as well as the angle (27 degrees) in the R.A., Decl. frame are shown. The three TDGs are in the same tidal tail originating from the outskirts of one of the two progenitors.
              }
         \label{Model6B-HR}
   \end{figure}
%

\section{Formation of NGC 1052 through a 1:1 merger model}
\label{sec:merger}
N-body hydrodynamical simulations of major merger remnants have successfully reproduced giant spiral galaxies (NGC 5907 and NGC 4013, \citealt{Wang2012,Wang2015}), including their kinematics and their very faint loop systems. Similar attempts have been made to reproduce the M31 galaxy and its haunted halo \citep{Hammer2018} as well as its kinematics and its rotation curve \citep{Hammer2025}. More generally, it is expected that gas-rich progenitors of $\ge$ 1:4 mergers occur in sufficiently large numbers at $z$$\sim$ 0.65 in order to explain the evolution of the galaxy morphological sequence \citep{Hammer2009,Hopkins2009,Delgado2010,Sauvaget2018}. 

The formation of elliptical galaxies requires a different scheme because in their case, the goal is to minimise the residual angular momentum in the merger remnant. This minimisation has been successfully realised by \citet{Wang2020}. Because the latter simulation provides very elongated tidal tails, we decided to adopt their Model-6 (see Table 1 of \citealt{Wang2020}; the model's initial gas-to-baryon fraction is 20\%) with very few adaptations to generate Model-6a. In particular, we reduced the mass of each particle by a factor of two because the stellar mass of NGC 1052 ($\sim$ $10^{11}$ $M_{\odot}$, \citealt{Forbes2019}) is approximately half that of Cen A, and we thus slightly reduced the initial fraction of DM  to 80\% (instead of 90\%) to align our modelling to that of M31. Our modelling is intended to be indicative of the structures formed during the major merger that has occurred in the history of NGC 1052. It does not intend to present an accurate reproduction of this galaxy (see Section~\ref{sec:discussion}). To adapt our simulations\footnote{Simulations have been carried out with the GIZMO code \citep{Hopkins2015}, which is based on a new Lagrangian method for hydrodynamics. In the code we implemented star formation and feedback processes as described in \citet{Wang2012}, following the method of \citet{Cox2006}. The cooling process was implemented in GIZMO with an updated version of \citet{Katz1996}.} to the ages of the DM-deficient dwarfs (8.9 Gyr for DF2 and DF4, \citealt{Fensch2019a}), we considered simulation durations of more than 10 Gyr after the merger conclusion.

As in \citet{Wang2020}, the 1:1 merger remnant is an elongated elliptical galaxy quite similar to an E4 galaxy, such as NGC 1052 \citep{Forbes2019}. Figure~\ref{Model6B-HR} shows the snapshot 128 that corresponds to 8.4 Gyr after the merger accomplishment. We chose this snapshot because the remnant halo shows a straight stellar stream on the left and a curved stream on the opposite side, which show similarities with the two main features discovered by \citet[see their Fig. 1]{Muller2019} in the NGC 1052 halo. These streams are associated with stellar particles coming back from the tidal tails that form loops after orbiting around the galaxy remnant. The streams form in a way similar to the long-lived loops observed in NGC 5907 and reproduced by \citet{Wang2012} 8.5 Gyr after the merger. The observed straight stellar stream of \citet{Muller2019} likely results from a loop seen edge-on, and to reproduce its narrowness would require a better coherence for the simulated stream, which could be optimised following the procedure described in \citet{Amorisco2015}. 
In Appendix~\ref{AppendixA} we present a slightly modified model that accounts for the HI gas and its misalignment with the NGC 1052 stellar body.

\section{Reproduction of the main observed properties in the NGC 1052 field}
\label{sec:reproduction}
Figure~\ref{Model6B-HR} shows a remnant elliptical galaxy 8.4 Gyr after the merger with tidal features and with TDGs in the field. As in \citet{Wang2020}, two almost aligned tidal tails are generated, with a total extent that can reach $\sim$ 1 Mpc. Here, we have chosen to present a configuration that may resemble NGC 1052 and its accompanying dwarfs, namely DF2, DF4, and DF7, to illustrate that a tail may encompass the whole field of view discussed in \citet{Keim2025}, which is larger than the one-degree field observed by \citet{Muller2019}. 

To verify whether DM-deficient dwarfs DF2 and DF4 can be reproduced by TDGs, we chose a model snapshot for which the alignment approximately resembles the observed NGC 1052 field. Since it is based on a model made to reproduce another elliptical galaxy (Cen A), it was quite surprising to retrieve approximately similar structures to what is observed in the NGC 1052 field. Table~\ref{tab:TDGs} provides a comparison between the TDGs shown in the inserts of Figure~\ref{Model6B-HR} with properties of three dwarfs in the NGC 1052 field. Velocities relative to NGC 1052 behave similarly to those observed, although with a smaller amplitude. Stellar masses of the simulated TDGs are smaller than those of the observed dwarfs, while they share their observed large extent. In Appendix~\ref{AppendixA} we present another merger we studied (Model6-b) with a similar orbit but a smaller pericentre, and the simulated TDGs are more massive than the observed dwarfs (see TDGXb in Table~\ref{tab:TDGs}). In summary, we made modest efforts to simulate a trail of DM-deficient dwarfs with quite similar geometry and properties to what is observed. We also notice that TDG3, which lies near the tip-end of the tidal tail, shares similar discrepancies in velocity and in position to DF7 since both are offset by the trail of dwarfs discovered in the surroundings of the DF2--DF4 line, although with different amplitudes. 

\section{Discussion}
\label{sec:discussion}
We verified that the properties of the DM-deficient dwarfs in the NGC 1052 field can be explained by TDGs populating tidal tails coming from the major merger resulting in a giant elliptical galaxy similar to NGC 1052. Here, we compare this scenario with that of the bullet-dwarf promoted by \citet{Silk2019}. We notice that the two scenarios have similarities: (1) For both kinds of progenitors, the collision produces tidal material that is DM-free (i.e. TDGs), and (2) the collision of the two progenitor dwarfs as well as that of the two galaxy progenitors of NGC 1052 should be head-on. Despite this, and for clarity reasons, we continue to refer to both alternatives as the TDG and the bullet-dwarf scenarios, respectively.  

\begin{table}[!htb]
\begin{center}
\caption{ Velocities, velocity dispersions, stellar masses, and half-light radii of the modelled TDGs compared to observed DF galaxies. }
\label{tab:TDGs}
\begin{tabular}{lrrrrl}
\hline
Name & $v_{\rm los}$ & $\sigma_{\rm vlos}$ & $M_{\rm star}$  & $r_{\rm half,2D}$ & Ref. \\
     & km~s$^{-1}$  & km~s$^{-1}$ & $10^7\,M_\odot$   & kpc & \\
\hline
DF2 & 245 & 6.1 & 20 & 2.2 &  1,3\\
TDG1a & 168  & 4.0 & 1.13  & 1.58 & M6a \\
TDG1b & 166 & 11.7 & 17.2  & 1.41 & M6b \\
TDG1c & 159.9  & 15.0 & 24.3  & 1.20 & M6c \\
\hline
DF4 & -127  & 7.7 & 15 &1.6 & 2,3  \\
TDG2a & -2.9  & 2.4 & 0.58  & 1.76 & M6a \\
TDG2b & -39  & 10.8 & 12.9  & 2.06 & M6b \\
TDG2c & -63.4  & 9.5 & 8.12  & 1.71 & M6c \\
\hline
DF7 & 130 & -  &1 &1.24 & 3,4,5 \\
TDG3a & 10.2  & 2.0 & 0.52  & 1.77 & M6 \\
TDG3b & -38 & 12.1 & 18.6  & 1.83 & M6b \\
TDG3c & -68.8  & 10.2 & 14.0  & 1.97 & M6c \\
\hline
\end{tabular}
\tablefoot{
Radial velocities are given relative to NGC 1052, for which a radial velocity of 1560 km~$\rm s^{-1}$ \citep{Koss2022} has been adopted. References are 1,2=\citet{vanDokkum2018}, \citet{vanDokkum2018}; 3= \citet{Keim2025}; 4= \citet{Tang2025}; and 5=\cite{Roman2021}, respectively. Properties of the M6a,b,c models are given in Table~\ref{tab:merger_models}.}
\end{center}
\end{table}

The estimated 3D distance of 1.7$\pm$0.5 Mpc  between DF2 and DF4 \citep{Shen2023} would mean that the trail of dwarfs extends almost exclusively along the line of sight. This estimate is essential for the bullet-dwarf scenario because it is the only way to account for an ancient-dwarf collision consistent with the DF2 age, while it could be fatal for the TDG scenario, which predicts a tidal tail elongated along all directions but limited to $\sim$ 1 Mpc in length. During the refereeing process, a new study by \citet{Tang2026} has changed the TRGB distance estimate of DF2 from 21.2$\pm$1.7 \citep{Shen2023} to 17.6$\pm$0.6 Mpc, the latter value being based on robust JWST observations. Such a considerable change presents serious doubts about the possibility of determining the radial distance from DF2 to DF4 using the former Hubble Space Telescope measurements. According to \citet{Tang2026} distance estimates from surface-brightness fluctuations, DF2 and DF4 could have the same distance within less than 1$\sigma$. However, much deeper photometric data from JWST are necessary for a robust estimate of the distance between DF2 and DF4.

In this work, we did not search for a precise reproduction of the observations and only sought to establish whether or not a TDG scenario could be plausible.\footnote{During the refereeing process, we were aware of the study by \citet{Bellstedt2018} of the NGC 1052 velocity maps. Our model reproduces well the amplitude of the velocity and velocity dispersion maps. However, it did predict an inversion of the velocity gradient due to our initial choice of adopting the \citet{Wang2020} modelling. This choice would not affect the plausibility of the TDG scenario because another velocity field orientation still produces tidal tails and TDGs, and it would require another way to reproduce the dwarf spatial and velocity properties.} This is because (1) the distances of all the concerned galaxies require TRGB measurements with JWST; (2) some features such as TDG number, positions, and velocities vary temporally; (3) there is an absence of knowledge about the NGC 1052 merger and star formation histories; and (4) the parameter space is much larger than that of the collision of two dwarfs, for which \citet{Lee2024} tested more than 1.5 million simulations. The three models (see Table~\ref{tab:merger_models}) shown in this work are of the same family; i.e. we changed very few parameters from one model to another. One may be surprised that the small exploration of parameters presented here provides results as consistent as those of \citet{Lee2024} for reproducing DM-deficient dwarf properties as well as their potential alignment. 

The TDGs produced in our modelling are faint and extended and have a low surface brightness, while \citet{Lee2024} instead recovered a compact DF2 and, for all but one case,\footnote{Simulations by \citet{Otaki2023} were able to form one 2.5 kpc-size dwarf amongst four realisations of sub-halo collisions, but the authors worked with a much smaller numerical resolution than that of \citet{Lee2024}.} a compact DF4. Depending on the model parameters, the recovered TDGs may have masses similar to those of DF2 and DF4. This is in agreement with observations \citep{Kaviraj2012} establishing that for a merger remnant mass similar to that of M31 or NGC 1052, the TDG stellar mass can reach 3 $\times$$10^{9}$ $M_{\odot}$ \citep{Fouquet2012}. 

 We were not able to reproduce the population of GCs surrounding DF2 and DF4 because of a technical resolution issue.\footnote{\citet{Lee2024}  dwarf progenitors are $\sim$ 20 times less massive than our giant galaxy progenitors. To be able to form $10^{6}$ $M_{\odot}$ GCs, \citet{Lee2024} required using $\ge$$10^{7}$ particles, while we would need $\ge$ 2$\times$ $10^{8}$ particles, a very uncomfortable number for hydrodynamical simulations.} However, mergers of galaxies are accompanied by a period of enhanced star formation. This process can lead to the formation of star clusters, some of which may survive as GCs \citep{Valenzuela2024} if they remain outside the dense regions, for example, if they are transported in a tidal tail. The over-luminous dwarf GCs appear to support the bullet-dwarf scenario.\footnote{However, we notice that DF2, DF4 and the two DM-deficient dwarf galaxies discovered by \citet{Buzzo2026} have been identified from GC kinematics. Detection of DM-deficient dwarfs could be biased towards detecting exceptionally bright GC systems, while \citet{Keim2026} identified DF9 as DM-deficient from its stellar velocity dispersion since it possesses a single central GC. Thus, the bullet-dwarf scenario is favoured, but one cannot exclude a specific TDG scenario providing over-luminous GCs in some DM-deficient dwarfs.}

Table~\ref{tab:TDGs} shows that the velocity of DF2 and DF4 relative to NGC 1052 is smaller than that observed while showing the same direction. A more precise match could be done by ignoring the presence of TDGs and simply changing their orientation in the radial direction, given the $>$ 1 Mpc size of the overall tail system of our model (see Figure~\ref{alignment} in Appendix~\ref{AppendixA},  Model-6c). We may also reproduce the velocity offset of NGC 1052 when compared to the line joining DF2 and DF4 velocities (see Fig.3 of  \citealt{Keim2025}). This offset has been argued to disfavour an association of the observed trail of dwarfs with NGC 1052, which is essential for the TDG scenario. However, the offset is explained by our modelling because TDG1 (alias DF2), which is the closest dwarf in projection to NGC 1052, is coming back from a tidal tail and starting to be captured into an orbit around the giant elliptical. Perhaps the main argument against a TDG interpretation is related to their supposed metal richness, while DF2 and DF4 are metal poor. However, this cannot apply if NGC 1052 is a several gigayear-old merger that would form TDGs inheriting the metal abundance of the outer disk of the merger progenitor, which could be low at high redshift \citep{Recchi2015}.

Both scenarios predict progenitor remnant(s), which is NGC 1052 for the TDG scenario. For the bullet-dwarf scenario, RCP 32 and DF7 have been invoked by \citet{vanDokkum2022} to be the progenitor dwarf remnants. However, they are reproduced in the \citet[see the bottom row of their Figure 5]{Lee2024} simulations with stellar surface brightnesses similar to that of DF2 or DF4. In particular, RCP32 has a central surface brightness three magnitudes fainter than that of DF2 or DF4 \citep{Roman2021}. Reproducing it as a bullet-dwarf progenitor would be an astonishing result because it is difficult to figure out how a dwarf can keep its DM content but lose all its baryons after a very violent collision. 

\section{Conclusions}
\label{sec:conclusion}
In this work, we have verified that a major merger experienced by NGC 1052 produces tidal tails and TDGs that match the intrinsic properties of DM-deficient dwarfs discovered by \citet{vanDokkum2018,vanDokkum2019} as well as their alignment along a trail. Our finding results from simple adaptations of the N-body hydrodynamical simulations made by \citet{Wang2020} to reproduce the properties of Cen A, and we addressed the supposed failures of the TDG scenario. The only limitation of our modelling comes from its very large parameter space and insufficiently precise constraints on NGC 1052 and associated dwarfs, which prevented us from experimenting with a more accurate reproduction of the observations. 

According to the hierarchical scenario, the present-day E4 morphology of NGC 1052 takes its origin from its last major merger, which requires radial orbits, moderate gas fraction, a mass ratio close to 1:1, and specific orientations of the progenitor galaxies \citep{Wang2020}. The latter conditions are necessary to remove most of the angular momentum in the remnant, which is currently accompanied by large tidal tails and TDGs and match better the observations when compared to the bullet-dwarf scenario. 

The bullet-dwarf scenario suffers from two major difficulties: first, reproducing the dwarf progenitors and, second, providing far too compact DM-deficient dwarfs. The scenario also neglects the fact that NGC 1052 is an active nucleus E4 galaxy showing tides, and it has likely experienced a major merger that can produce TDGs resembling the observed DM-deficient dwarfs. The likely occurrence of massive galaxy mergers near NGC 1052 has to be compared to the very rare dwarf--dwarf collision events in the field, for which only ten have been found in a $\rm (100 Mpc)^3$ volume from $z$=3 to $z$=0 \citep{Lee2024}. Future observations of dwarf radial distances will probe which of the two scenarios is the most likely. The dwarf GC over-luminosity appears to support the bullet-dwarf scenario, though there might be subtle selection effects associated with this property. 

The above makes  plausible an association of DM-deficient dwarfs and their trail with TDGs lying in a tidal tail formed during a major merger, whose remnant is consistent with the E4 morphology of NGC 1052. A better estimate of when NGC 1052 experienced its last merger would be welcomed, as it is currenlty unconstrained due to a lack of observations to establish its star formation history.  This paper demonstrates that the numerous mergers that massive galaxies have experienced in their past may have also formed TDGs \citep{Barnes1992,Okazaki2000}. The detection of DM-deficient dwarfs in the NGC 1052 field may confirm that TDGs can survive 8 Gyr after their formation, and they may represent a non-negligible part of the dwarf population (see also \citealt{Pawlowski2014}).
\begin{acknowledgements}
We are grateful to Jianling Wang for suggestions and providing us the initial conditions of his Model 6, to Piercarlo Bonifacio for enlightening discussions and advices about the distance determination  from TRGB, and to Istiak Akib, Marcel Pawlowski and Joe Silk for discussions and suggestions. We warmly thank the referee whose suggestions and comments help us to considerably improve the manuscript. This work was granted access to the HPC resources of MesoPSL (reference
ANR-10-EQPX-29-01).
      \end{acknowledgements}

%
\bibliographystyle{aa} 
\bibliography{NGC1052.bib} 
%
%
%

\begin{appendix} 
\section{Test of more radial orbits}
\label{AppendixA}
\subsection{Reproducing the residual HI gas in the NGC 1052 centre}

The NGC 1052 centre shows a fast rotating ($V_\mathrm{rot}$= 200 km$\rm s^{-1}$) gas component  \citep{vanGorkom1986}, which represents less than one percent of its baryonic mass. Such a component is likely responsible for the tiny recent star formation in the very inner part of NGC 1052, with ages varying from 1 to 4 Gyr \citep{Fernandez-Ontiveros2011}. Perhaps surprisingly, 8.4 Gyr after the merger occurrence, the simulated galaxy still shows the presence of rotating gas and of recent star formation in the centre. Many elliptical galaxies present residual HI gas in their centre, and similar to Cen A, NGC 1052 shows a gas-disk orientation that is almost perpendicular to the optical major axis \citep{vanGorkom1986}. \citet{Wang2020} was able to reproduce a similar behaviour for Cen A, for a shorter post-merger epoch than for NGC 1052. Recent simulations \citep{Porter2026} have shown that the gas component may survive and star formation may be mostly prevented because of Coriolis forces that allow the gas component to precess. 

\begin{figure*}
   \centering
\includegraphics[width=13cm]{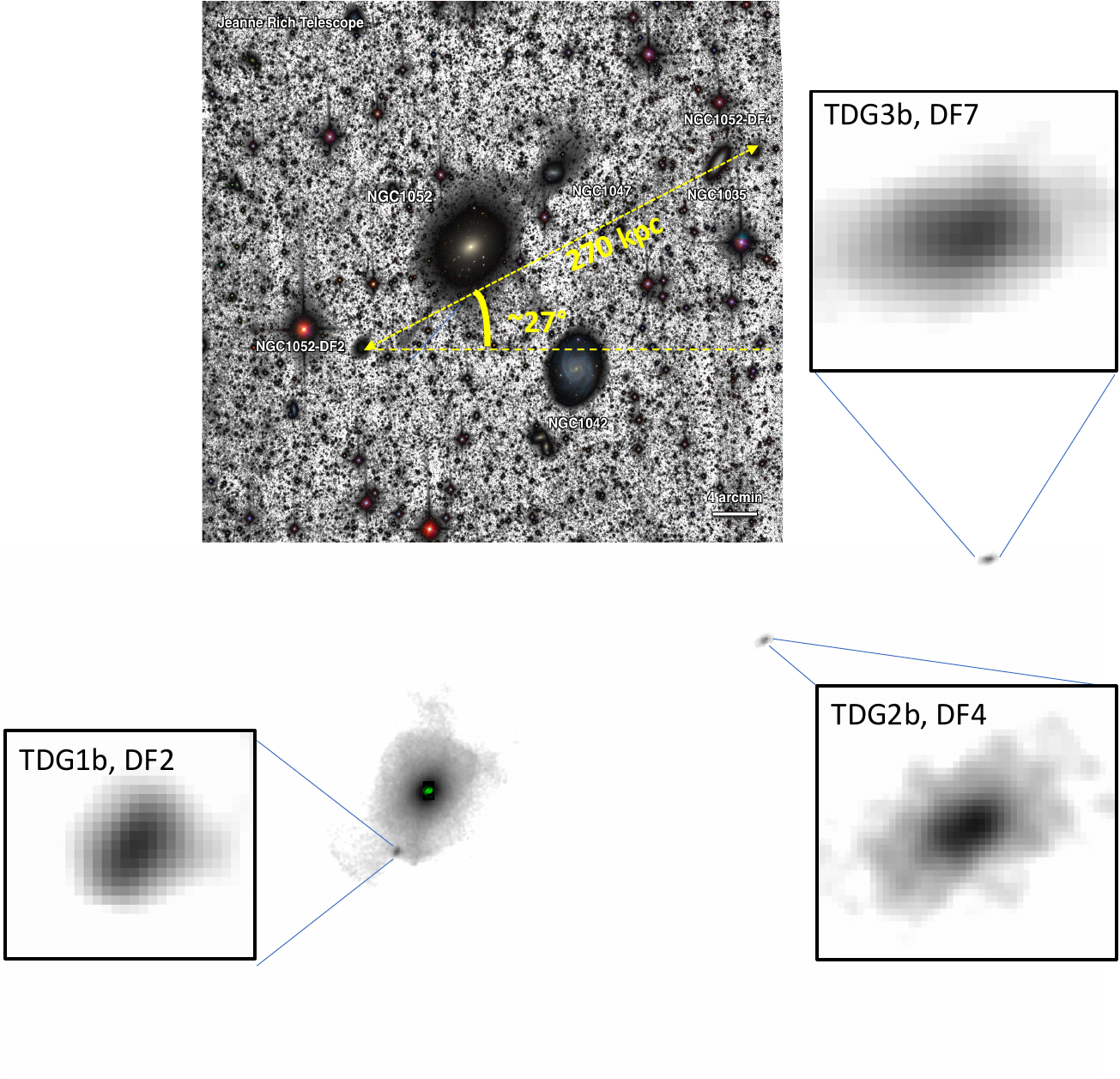}
      \caption{Comparison between the deep observations by \citet[see top-left image]{Muller2019} and \citet{Wang2020} adapted simulations for stellar particles (Model-6b). Three TDGs are zoomed in on the insert panels and they may resemble DF2, DF4 and DF7 (out of the observed field of view), while their properties are compared in Table~\ref{tab:TDGs}. The spatial alignment shows many similarities with observations. NGC 1052 is reproduced as an E4 galaxy with tidal features observed by \citet[see top-left image]{Muller2019}, but the halo orientation is offset by 20 degrees. The green colour image at the centre represents the gas disk that is almost perpendicular to the main axis of the E4 galaxy, though with a much smaller extent than that observed by \citet{vanGorkom1986}. 
              }
         \label{Model6B-80_rp7_08}
   \end{figure*}

To test whether the gas rotating disk can be found on an axis almost perpendicular to the simulated stellar ellipsoid that represents NGC 1052, we explored a range of merger orbital parameters (pericentres and inclination of progenitor galaxies), requiring the model to also reproduce the general properties of NGC 1052 and the alignment of the dwarf galaxies. Our goal has been to show the formation of a perpendicular residual gas disk as being strongly linked to the progenitor spins, favouring highly retrograde and polar orientations with respect to the orbital angular momentum. Here, we show one of such model (Model-6b) in which we have decreased the pericentre to 7 kpc (instead of 20 kpc), and by flipping the orientations of the two galaxies (see Table~\ref{tab:merger_models}). The first goal has been realised, i.e. the gas rotating  disk can be found on an axis almost perpendicular to the simulated NGC 1052, and interestingly, we have improved the TDG mass, velocities, and orientation of the NGC 1052 stellar halo (compare Figure~\ref{Model6B-80_rp7_08} to Figure~\ref{Model6B-HR}). Figure~\ref{Model6B-80_rp7_08} shows how this adapted model reproduces the observations, but at an epoch closer to the merger (4.7 Gyr instead of 8.4 Gyr). However, since few stars have been recently formed in such a gas-poor model, we have verified that most of the TDG stars are very old and consistent with the \citet{Fensch2019a} age of 8.9$\pm$ 1.8  Gyr. Moreover, this simulation provides TDGs with large stellar masses that are even larger than those of DF2, DF4 and DF7 (see Table~\ref{tab:TDGs}). This can be explained by the fact that highly radial orbits are accompanied by large tidal tails in order to remove most of the residual angular momentum. Notice also that TDG3b is very elongated as is observed for DF7 by \citet{Keim2025}.  We also notice that the shape of the HI emission may include some distorted features, which could be taken as an evidence of a more recent merger event. 
Besides this, we do not analyse the gas content of the TDGs, because  in a galaxy group such as that of NGC 1052 the ram pressure stripping provides a means of depleting the low-density gas reservoir (see \citealt{Silk2019}, and references therein).  \\

%
\subsection{Reproducing the DM-deficient dwarf alignment on sky}
\begin{figure*}
   \centering
\includegraphics[width=13cm]{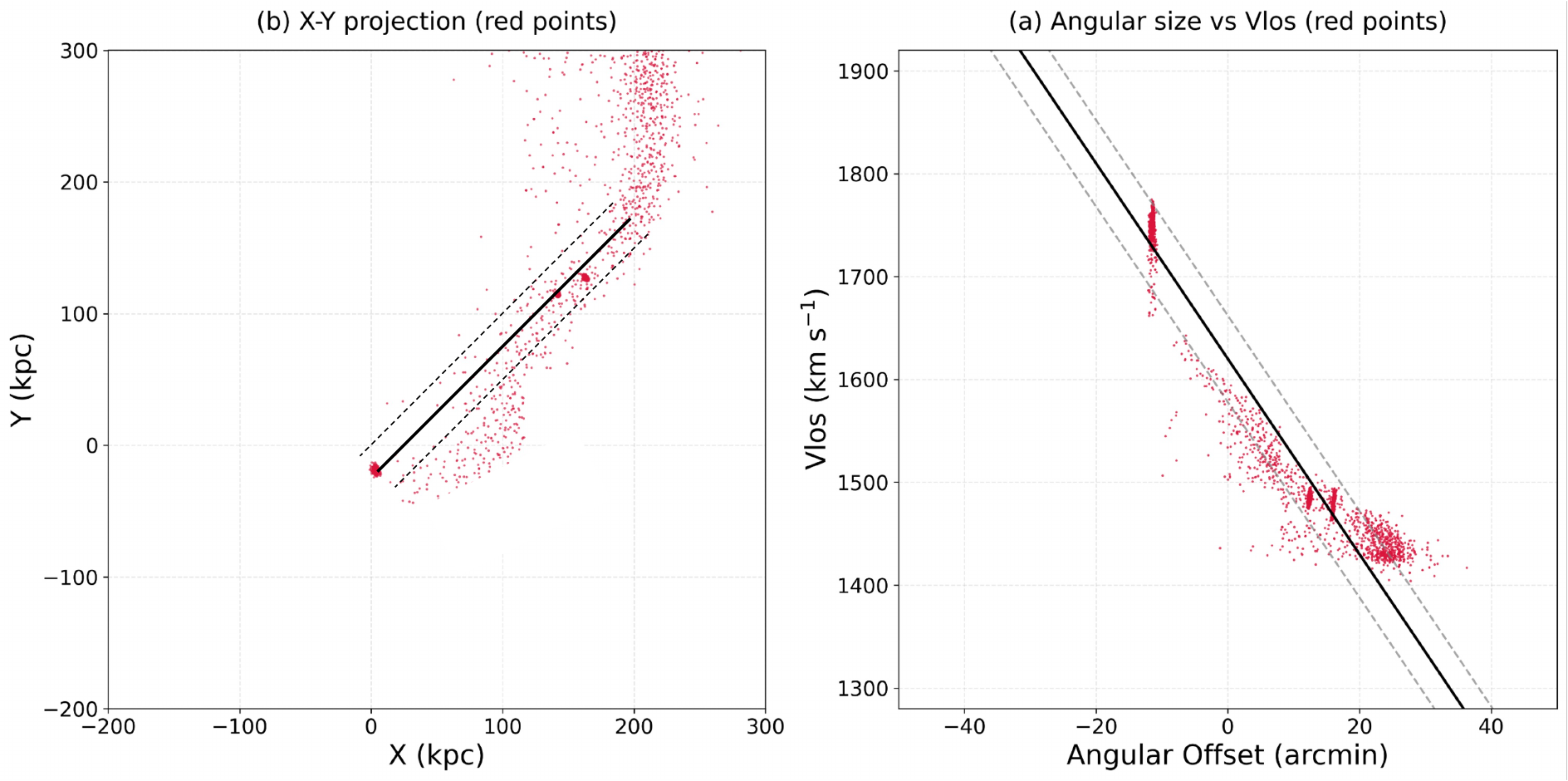}
      \caption{Reproduction of the alignment between three TDGs and several stellar particles (red dots) of the tidal tail after using the same model as Model-6b of Figure~\ref{Model6B-80_rp7_08}, but with a larger orbital eccentricity (Model6-c). Here, we do not search for reproducing the precise positions of the observed DM-deficient dwarfs because we are not able to fix their positions in the simulated tidal tails. (Left): Alignment in the sky over 270 kpc, where DF2 would be represented by the leftmost TDG, DF9 by one of the other TDGs, while DF4 would be at the right-end of the black line. (Right): Same as the left panel but showing the radial velocity versus the position along the DF2--DF4 axis.
              }
         \label{alignment}
   \end{figure*}

\citet[see their Fig. 3]{Keim2025} showed that the three DM-deficient dwarfs DF2, DF9 \citep{Keim2026} and DF4 are both spatially and kinematically aligned. We acknowledge that it could appear challenging to capture the alignments this way, given the curvature of most tidal tails in both space and velocity. However, its realisation progressively improves by enhancing the orbital eccentricity (then the initial energy) from 0.9 to 1.1, after using the same model than in Figure~\ref{Model6B-80_rp7_08}. Figure~\ref{alignment} shows that the three-dwarf alignment can be realised, although the constraint would be more stringent if RCP26 is found to be also a DM-deficient dwarf. The alignment is considered as a strong argument for the bullet-dwarf model because it automatically creates a line in space and in velocity due to the very radial orbit that has been adopted by, for example, \citet{Lee2024}. One difficulty in aligning TDGs in a tidal tail is especially due to the gravitational field of the remnant that gradually captures the most bound TDGs, and then modifies their radial velocity as it does  in the bullet-dwarf model, and finally the alignment is similarly reproduced for both alternatives (compare Figure~\ref{alignment} with Fig. 8 of \citealt{Lee2024}).

%

\subsection{Parameters of the modelling}
Table~\ref{tab:merger_models} presents the general properties of the three models used in this paper, illustrating that they are part of the same family with differences that only concern pericentres and the initial angles of the progenitors. 

\begin{table}
\centering
\caption{Initial parameters of the three merger models. }
\label{tab:merger_models}
\begin{tabular}{lccc}
\hline
Parameters & Model-6a & Model-6b & Model-6c \\
\hline
Mass ratio & 1.0 & 1.0 & 1.0 \\
Gal1 incx & 0 & 0.0 & 0.0 \\
Gal1 incy & 90 & 70.0 & 70.0 \\
Gal1 incz & $-90$ & $-90.0$ & $-90.0$ \\
Gal2 incx & 0 & 21.17 & 21.17 \\
Gal2 incy & 70 & $-18.75$ & $-18.75$ \\
Gal2 incz & $-40$ & $-82.90$ & $-82.90$ \\
Gal1 gas fraction & 0.2 & 0.2 & 0.2 \\
Gal2 gas fraction & 0.2 & 0.2 & 0.2 \\
Gal1 $h_{\rm star}$ (kpc) & 5.1 & 5.1 & 5.1 \\
Gal2 $h_{\rm star}$ (kpc) & 5.1 & 5.1 & 5.1 \\
Gal1 $h_{\rm gas}$ (kpc) & 10.2 & 10.2 & 10.2 \\
Gal2 $h_{\rm gas}$ (kpc) & 10.2 & 10.2 & 10.2 \\
$r_{\rm peri}$ (kpc) & 20 & 7 & 7 \\
Eccentricity & 1.0 & 1.0 & 1.11 \\
$N_{\rm particle}$ & 1.85M & 1.85M & 1.85M \\
$m_{\rm dm}:m_{\rm star}:m_{\rm gas}$ & 4:1:1 & 4:1:1 & 4:1:1 \\
\hline
\end{tabular}
\tablefoot{
For Model-6b and Model-6c, the listed disk orientation angles are the effective single-rotation angles equivalent to the second rotation matrix applied to Model-6.}

\end{table}

\end{appendix} 

\end{document}